# Analyzing the Persistence of Referenced Web Resources with Memento

Robert Sanderson
Los Alamos National Laboratory
Los Alamos
NM 87544, USA
+1 (505) 665-5804
rsanderson@lanl.gov

Mark Phillips
University of North Texas
Denton
TX 76203, USA
+1 (960) 565-2415
Mark.Phillips@unt.edu

Herbert Van de Sompel
Los Alamos National Laboratory
Los Alamos
NM 87544, USA
+1 (505) 667-1267
herbertv@lanl.gov

## ABSTRACT
In this paper we present the results of a study into the persistence and availability of web resources referenced from papers in scholarly repositories. Two repositories with different characteristics, arXiv and the UNT digital library, are studied to determine if the nature of the repository, or of its content, has a bearing on the availability of the web resources cited by that content. Memento makes it possible to automate discovery of archived resources and to consider the time between the publication of the research and the archiving of the referenced URLs. This automation allows us to process more than 160000 URLs, the largest known such study, and the repository metadata allows consideration of the results by discipline. The results are startling: 45% (66096) of the URLs referenced from arXiv still exist, but are not preserved for future generations, and 28% of resources referenced by UNT papers have been lost. Moving forwards, we provide some initial recommendations, including that repositories should publish URL lists extracted from papers that could be used as seeds for web archiving systems.

## Categories and Subject Descriptors
H.5.4 [**Information Interfaces and Presentation**]: Hypertext/Hypermedia – *Architectures, Navigation.*

## General Terms
Experimentation

## 1. INTRODUCTION
As repositories become more aligned with the web architecture and links to and from their content proliferate, the role of the repository moves away from that of a curated content silo and toward knowledge infrastructure for research. This infrastructure is the foundation of the entire research community, and with scholarly communication in the midst of the transition from print to digital, resource preservation has become an area of increasing concern.

The current generation of repositories performs the important task of preserving copies of scholarly research output, however maintaining access to research inputs is equally crucial to enable future understanding and reproducibility. Those inputs, both data and prior work, are increasingly online and often not maintained within the stable and managed confines of a repository.

This paper considers the extent to which web resources, referenced throughout academic works in repositories, are still available. Using the Memento HTTP extensions [15] for access to historical web content, we can go beyond previous studies and determine not only if there is still a resource at the cited URL, but whether or not there are copies in archives and even consider the difference between the publication date of the citing work and the time of the closest archived copy. The time difference is a crucial factor not previously considered, as the information at any given URL can be modified at any time. A copy is thus more likely to be an accurate representation of the intended target of the citation if it is archived close to the publication date of the paper.

This research utilizes two repositories of scholarly communication of very different types. The first is an institutional repository of the electronic theses and dissertations of the students of the University of North Texas[1]. Although the total number of documents is relatively low, less than 4000, the subject matter is wide ranging, covering the full spectrum of a modern university. The second collection analyzed, arXiv[2], is a large multi-institutional subject repository of published works in High Energy Physics and related disciplines. The number of documents is two orders of magnitude higher; in the order of 400000.

By processing the referenced URLs and metadata from works stored in the two repositories, we aim to address the questions:

1. To what extent are web resources referenced from works in the repositories still available at their original URLs or from archives of web resources?
2. How long is the period between the publication of a paper and the archiving of a resource cited by the paper?
3. Does the nature of the repository or the academic level and discipline of the work have any bearing on the previous two questions?

The answers to these questions will determine if there is a need for repositories to take an active role in the preservation of referenced web resources and the natures of that role.

## 2. BACKGROUND WORK
Much research has been invested already in analyzing "link rot" (when a hyperlink fails to resolve) in academic publications across various disciplines. These studies give us a baseline expectation for the availability of resources at their original URLs, but the sample size is often very small. Previously, the number of URLs checked to ascertain if the content existed elsewhere was also limited by the manual nature of such checks.

---

[1] http://digital.library.unt.edu/

[2] http://www.arxiv.org/

Table 1 summarizes the results of 17 such studies, showing the number of URLs considered, the period in which the referencing works were published, and how many were later available. While the most studied discipline is Information Science, Rumsey [13] has looked at Law, Wren [18,19], Falagas [7] and Wagner [17] look at Biology, Dimitrova [4] at Communication publications, Russell [14] at History journals, Duda [5] at Ecology, Parker [12] studies New Zealand journals, and Davis [3] looks at undergraduate papers from an Economics course.

| Paper | URLs | Years | Available |
|---|---|---|---|
| Casserly 2003 | 500 | 1999-2000 | 56.4% in 2002 |
| Casserly 2007 | 500 | 1999-2000 | 39% in 2006 |
| Russell 2008 | 510 | 1999-2006 | 82% in 2006 |
| Davis 2002 | 688 | 1999-2001 | 69% in 2002 |
| Sellitto 2005 | 1043 | 1995-2003 | 54% in 2005 |
| Dimitrova 2007 | 1126 | 2000-2003 | 61.3% in 2007 |
| Parker 2007 | 1229 | 2002-2005 | 70% in 2007 |
| Falagas 2007 | 1417 | 2003-2006 | 83% in 2006 |
| Wren 2004 | 1630 | 1994-2002 | 63% in 2003 |
| Moghaddam 2010 | 1761 | 1995-2008 | 73% in 2010 |
| Wagner 2009 | 2011 | 2002-2004 | 50.7% in 2007 |
| Duda 2008 | 2100 | 1997-2005 | 70-80% in 2008 |
| Goh 2005 | 2516 | 1997-2003 | 69% in 2005 |
| Rumsey 2002 | 3406 | 1997-2001 | 48% in 2001 |
| McCown 2005 | 4387 | 1995-2004 | 70% in 2005 |
| Wren 2008 | 6154 | 1994-2007 | 80% in 2008 |
| Lawrence 2001 | 67577 | 1993-1999 | 75% in 2000 |

**Table 1. Summary of Previous Studies**

Of the papers summarized, several attempted to discover if the resources that were not available at their original URL were still online at new locations. Lawrence [9] used search engines to investigate the availability of 205 URLs that did not resolve and rediscovered 163 (79.7%). Moghaddam [11] increased their availability rate from 73% to 86% using Google. Casserley and Bird in 2003 [1] used searches to improve their rate to 81.4%, and up to 89.2% by also searching the Internet Archive. In Casserley 2007 [2], they again improve from 39% to 73% via searches, and up to 81% using the Internet Archive. Falagas [7] increases from 83% to 97% using Google. Wagner [17] finds 593 of 992 inactive URLs in the Internet Archive, and Russell [14] finds 57% of his missing links. Duda [5] finds 72-84% of the missing Ecology links using a search engine.

The use of search engines and the Internet Archive in the studies was done by hand, lacking access to an API for automatically determining if a copy of the resource existed elsewhere online. This functionality is provided by the Memento framework, that allows a system to resolve the tuple {Original URL, Desired Time} into the URL of the copy, termed a Memento, closest-in-time to Desired Time across all known, public web archives.

The Memento framework provides interfaces for both interactive use and for machine processing. TimeGates are web resources intended for use by browsers, and given the desired time in a newly defined Accept-Datetime header, they will redirect the browser to the temporally closest known copy of a resource. TimeMaps, on the other hand, are manifests of all of the known copies of a particular resource and the times at which they were archived.

## 3. DESIGN AND IMPLEMENTATION

Given automated access to the information contained in Memento TimeMaps, the previous work can be extended to a new level. Instead of searching by hand, the entire process of checking the persistence of the Original URL and whether or not copies exist in an archive can be automated. This makes comprehensive experiments well beyond the scale of previous work possible.

In order to answer the research questions posed in the Introduction, it was necessary to extract the URLs from the full text of the papers in the two repositories and match these with the publication date and subject areas of the paper. This initial data was collected using the following methodology:

For each paper:

1. Extract the text of the PDF to XML using pdftohtml[3]
2. Extract the encoded links using the XPath: //a/@href
3. Extract links from the text using a regular expression[4]
4. Normalize and filter for mistakes in the text and incorrect extractions
5. Deduplicate the sets of links
6. Extract publication and subjects from metadata
7. For each extracted URL, add a tuple to the dataset: {URL, paper identifier, publication date, subjects}

The filtering and normalization process in step 4 was as follows:

1. Ensure ascii encoding, reducing Unicode U+223C (Tilde Operator) to regular tilde '~', and discard any irreducible strings
2. Add 'http://' to the beginning, if necessary
3. Add a trailing slash, if no path is given
4. Discard if the URL's top level domain (.com, .org etc) is not in the list of known top level domains
5. Remove default port, discard if the port is not numeric
6. Reduce the hostname to lowercase
7. Discard if the link is matched in a host/URL blacklist

The blacklist was necessary to exclude example URLs, internal URLs such as localhost or UNT library proxy URLs, automatically inserted URLs such as links to other arXiv papers and sibling repositories such as SLAC-SPIRES, as well as other repositories and archives such as dx.doi.org and the LANL arxiv mirror that are considered to be already within the repository community and thus part of the internal, managed content.

This process produced 17965 unique URLs, and 19186 {URL, paper} combinations from the 3595 papers in the UNT repository. These were divided amongst a total of 128 disciplines, of which the top 20 made up 75% of the {URL, paper} combinations, as each paper has exactly one discipline. The publication dates range from May 1999 through to August 2010.

From the 400144 papers processed from arXiv, 144087 unique URLs were extracted, with 287266 {URL, paper} combinations.

---

[3] http://pdftohtml.sourceforge.net/

[4] http://daringfireball.net/2010/07/improved_regex_for_matching_urls

Each paper can have zero or more subjects, and these were collapsed to the top level used in the arXiv hierarchy, such that 'cs.dl' (Computer Science/Digital Libraries) would be reduced to Computer Science. This produced a total of 658966 {URL, paper, subject} combinations across the 36 top-level subjects. The publication dates of the documents range from December 1993 through to December 2009.

Using a Memento Aggregator that covers nine archives, including the Internet Archive, Web Citation, the UK National Archive and the Library of Congress, the implemented system attempted to retrieve a TimeMap for each of the 162052 cited URLs. If a TimeMap could not be retrieved, the URL was marked as not being archived. At the same time, the system also attempted to dereference the URL to determine if the resource still existed. Any HTTP Status Code less than 400 was considered a success.

The information from the TimeMaps was then processed to find the Memento for each cited URL with a datetime closest to the date of publication of the paper.

## 4. RESULTS

The results differ significantly between the two repositories tested, when the archived versus still exist distinction is drawn. As shown in Figure 1, for 72% of the URLs tested from UNT, Mementos were available in archives, the resource still existed at its original location, or both. This is in line with previous studies. 54% (9880) of the URLs were available in one or more archives, leaving 28% (5073) that could not be automatically discovered.

The red color in Figures 1 and 2 represents the lost resources, and the bright green resources that still exist and are archived. The peach colored segments are resources at risk of disappearing, and the light green do not exist but are preserved in an archive.

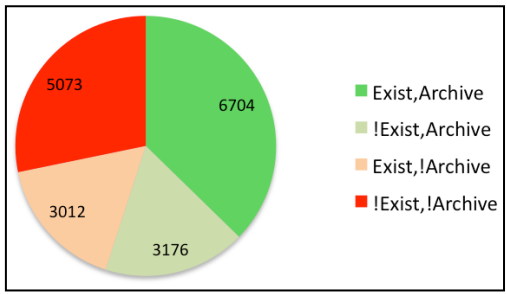

**Figure 1. Availability of URLs cited in UNT Repository**

The much larger arXiv dataset exhibited similar overall availability of 78%, however differed significantly from UNT otherwise. Only 32% were archived, and 70% still existed at their original location. The most dramatic finding is that 45% (66096) of the URLs that currently exist are not archived at all.

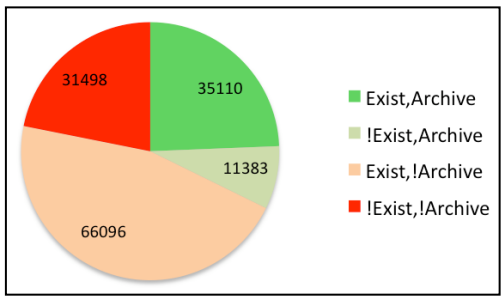

**Figure 2. Availability of URLs cited in arXiv**

When comparing discipline by discipline, the two repositories again exhibited different patterns. Some disciplines at UNT, depicted in Figure 3, such as Biology (177 URLs), Analytical Chemistry (131) and Behavior Analysis (134), had a very high percentage of their resources available as a Memento within a month of the publication of the paper. The largest by volume, such as Information Science (3310) and Educational Administration (1262) exhibited the same pattern of a slightly higher percentage archived and approximately half of that within a month of the publication. Some, such as History (435), Journalism (197) and English (154) exhibited a different pattern where more still exist and a very low percentage were archived within a month.

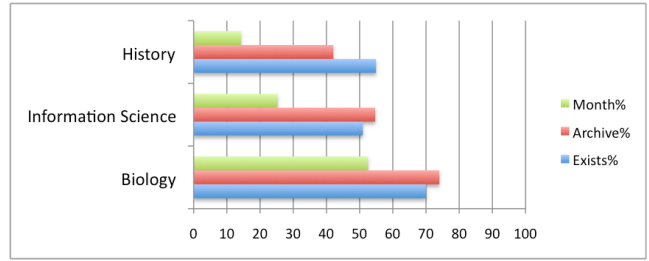

**Figure 3. Select Disciplines in UNT Repository**

The subjects in arXiv, Figure 4, all exhibit similar behavior of a very high percentage still in existence and a very low percentage archived within a month. The significant distinction is that Statistics and Mathematics have a much lower archival rate than the Physics categories and Computer Science. By volume, there are 83584 Astro-Physics URLs of which 66% are archived, yet only 39% of the 96592 Mathematics resources are preserved.

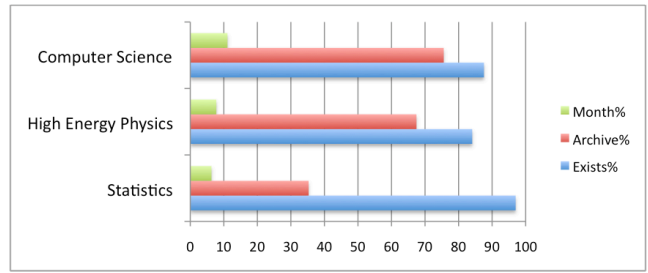

**Figure 4. Select Disciplines in arXiv**

The difference between publication date of a paper and the temporally closest archived copy is a typical long-tail graph. There are many URLs archived within a short amount of time, but a small number of URLs get archived a long time after the publication. Both UNT and arXiv exhibit extremely similar behavior, depicted with a log/log scale in Figure 5.

Of the archived Mementos, 48% were archived within one month of the publication of the citing paper, either before or after, for UNT and 45% for arXiv. 80% of the cited Mementos had been archived within a year of the publication for both repositories, the only difference is one of scale of total referenced resources. The average number of days for UNT was 224, and 217 for arXiv.

This means that even if there is a copy of a cited resource preserved in an archive, it is still likely not to be the representation served at the publication time of the citing work. Better preservation practices are needed in order to assess the impact of this time difference, and the true likelihood that the cited state of the resource is the one that is archived.

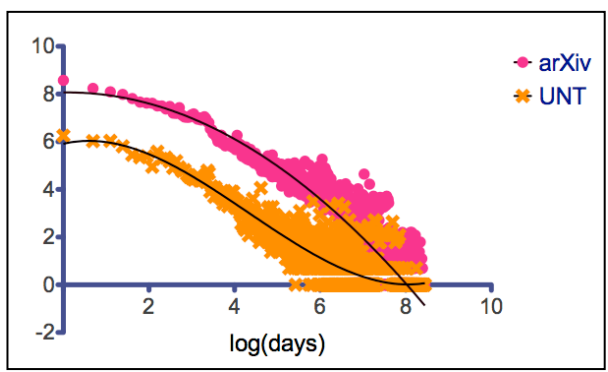

**Figure 5. log() of publication/archive date difference vs log() of count of such URLs**

## 5. CONCLUSIONS

There is a real need for repositories to become involved in the preservation of the scholarly communication record, beyond their own deposited content or other managed resources. As shown, a large number of referenced resources are at risk of disappearing. Our results suggest that as the disciplines get more technical, and the level of the research increases from student to professional, the proportion of archived cited material decreases significantly.

In the spirit of community and collaborative efforts, our recommendation is that repositories should expose the list of URLs referenced in each paper using an API such as an ATOM feed or other technology appropriate for the repository's platform. This feed could then be used as the seed list for web crawlers to ensure that the referenced resources are preserved. The repository community could also run a web crawler and archive to guarantee the persistence rather than relying on the goodwill of external agencies such as the Internet Archive. Using Memento, archives of this nature would be instantly accessible to readers who, as a result, would be able to faithfully reconstruct the context in which the paper was published.

While it is important that archives such as WebCite [6] exist that allow the preservation of the correct representation of cited resources, it is insufficient to rely on the authors to use such manual systems as the burden is similar to the extra work of filling out repository metadata forms. An automated and distributed system would have a significantly better chance at completeness and success.

This extraction by the repository would also make it easier for future research to occur on this topic, as the cooperation of the repositories in providing the full text of their collections was essential for this paper. Future work should verify the discoveries presented here by applying the same methodology to other repositories, further disciplines and different collection sizes. The authors intend to look at Citeseer and JSTOR, but would welcome further discussion and collaboration. Assessment of the rate of change of cited resources is also important to establish the maximum appropriate time between publication and archiving.

## 6. ACKNOWLEDGMENTS

This research was funded in part by the Library of Congress.